\documentstyle[12pt]{article}

\def\be{\begin{equation}}
\def\ee{\end{equation}}
\def\ba{\begin{array}}
\def\ea{\end{array}}
\def\bd{\begin{description}}
\def\ed{\end{description}}
\def\ket#1{ | #1 \rangle }
\def\setC{ \rm \rule[0.1ex]{0.07ex}{0.6em}\hspace{-0.27em}{C} }

\newfont{\hbb}{msbm10 scaled \magstep1}
\def\setS{\mbox{\hbb S}}

\def\titel{The Quantum Theory of Ur-Objects as a Theory of Information}
\begin{document}
\sloppy

\thispagestyle{empty}

\begin{center}

\vspace*{2ex}

{\Large \titel}\footnote{Published:
{\em International Journal of Theoretical Physics}, Vol. 34, No. 8,
p. 1541 - 1552, 1995
}

\vspace*{2ex}

{\large Holger Lyre}\footnote{
Institute of Philosophy,
Ruhr-University Bochum,
D-44780 Bochum,
FRG,\\
email: holger.lyre@rz.ruhr-uni-bochum.de
}

\vspace*{2ex}

{\large August 1994}

\vspace*{2ex}

\begin{abstract}
The quantum theory of ur-objects proposed by C. F. von Weizs\"acker
has to be interpreted as a quantum theory of information.
Ur-objects, or urs, are thought to be the simplest objects
in quantum theory. Thus an ur is represented by a two-dimensional
Hilbert space with the universal symmetry group $SU(2)$,
and can only be characterized as {\em one bit of potential information}.
In this sense it is not a spatial but an {\em information atom}.
The physical structure of the ur theory is reviewed,
and the philosophical consequences of its interpretation as an
information theory are demonstrated by means of some important
concepts of physics such as time, space, entropy, energy, and matter,
which in ur theory appear to be directly connected with information
as ''the'' fundamental substance. This hopefully will help to
provide a new understanding of the concept of information.

\begin{description}
\item[Keywords:] information, quantum theory, ur-object, symmetry group,
                 physical concepts
\item[AMS Classification:] 81P05, 81R99, 94A17
\end{description}
\end{abstract}

\end{center}


\section{Introduction}

This paper deals with a certain kind of quantum theory - the so-called
quantum theory of ur objects developed by C.~F.~von~Weizs\"acker and his
collaborators (Castell, Drieschner, G\"ornitz, et al.).
The ur theory can be regarded as a quantum theory of information.
The basic concepts of physics such as time and space are related
to the concept of information and classical physical substances
such as energy and matter, of which the world consists and
which could be regarded as equivalent since special relativity theory,
are reduced to information as ''the'' fundamental substance.
A short outline is given of the concept of space related
to information and to the connection between information on one hand
and energy and matter on the other.

\section{Space as a Representation of Information}

\subsection{Urhypothesis and the Concept of Position Space}

In ur theory one starts with the elementary assumption that any
object which in quantum theory is represented by a Hilbert space
spanned by the states of the object
can be described in a state space which is isomorphic
to a subspace of tensor products of two-dimensional complex spaces.
In a more logical formulation this means that the set of $n$
attributes or properties which are necessary to describe
a physical object in terms of its possible states
can be regarded as an $n$-fold alternative.
In ur theory any alternative will be decomposed into the
Cartesian product of elementary binary alternatives - called
ur alternatives or ur objects (urs).
This leads to a ''logical atomism'', i.e. the
smallest objects in physics are not small as regards their spatial 
but their logical smallness.
Thus an ur object can conceptually only be characterized as representing
{\em one bit of potential information}. In this sense ur theory basically
has to be understood as a quantum theory of information
and so information acquires a new dimension as the fundamental
physical substance.

We repeat the basic postulates of ur theory in a more formal
way
\bd
\item[Definition ''ur object'':] An ur is described by a twodimensional
     complex state vector
     \be
     \ket{ u_r } \in \ \setC^2 , \qquad r=1,2
     \ee
\item[Rule of State Spaces:] Hilbert spaces of any objects can be
     represented in a subspace of the tensor product space of
     two-dimensional Hilbert spaces belonging to urs
     \be
     \label{tensorspace}
     V^m \subseteq T_n = \bigotimes_{n} \ \setC^2 , \qquad m \le 2^n
     \ee
\item[Symmetry Group:] The universal symmetry group $Q$ of an ur object
     keeps invariant the unitary norm
     $\langle u | u \rangle = u_{1}^{*} u_{1} + u_{2}^{*} u_{2}$
     and contains the subgroups
     \be
     \label{Q}
     SU(2) \times U(1) \quad {\rm and} \quad K .
     \ee
     The antilinear transformations $\hat K \in K$ act like
     $\hat K \ket{u} = i \hat\sigma_{2} \ket{u^{*}}$, where
     $\hat\sigma_{2}$ is the second Pauli matrix
     and $^*$ the complex conjugation.
\ed
In ur theory the {\em three-dimensional position space} is derived as a
consequence of these mathematical conditions. This can be explained
by analyzing the concept of space. In most cases the spatial distance
between objects can be understood as the parameter for the interaction
between these objects. On the other hand, the definition of a physical
object (e.g., a massive elementary particle) depends on the separation of
its typical spatial range.
Supposing that all objects consist of urs, the total state of the
universe should remain unchanged by transforming all urs with the
same element from the symmetry group of the ur,
which is essentially $SU(2)$.
Thus the interaction between all objects should be invariant and
therefore the position space as a parameter space for the strength
of interaction should have the same structure as the symmetric space
of the symmetry group of the ur.
In ur theory therefore the assumption is made
that the position space has to be {\em identified}
with the homogeneous space $\setS^3$ of the group $SU(2)$.
Later the time development of urs will be described by
the group of phase transformations $U(1)$ in (\ref{Q}).

\subsection{Large Numbers in Physics}

The quantum theory of urs gives an argument for deriving the physical
position space and for describing its global structure as a space of
constant curvature $k=1$, i.e., a model for an Einstein cosmos.
Space in this sense appears as a {\em representation or realization
of information as the physical substance}.
In this connection it is useful to remark that we call only those binary
alternatives ur alternatives which lead back to {\em spatial} decisions,
i.e. decisions which can only be made in position space
(e.g., consider the spin state of an electron: its measurement by using
a Stern-Gerlach apparatus will be done by deciding a spatial alternative
about the deflection of the electron in an inhomogeneous magnetic field
and could therefore be looked upon as the decision of an ur alternative).
By using the central assumption in ur theory that all physical objects
consist of urs, it follows that all physical attributes or properties
of objects, insofar as we are able to decide them empirically,
are only measurable in position space.
Thus ur theory explains a general and indeed well-known experience
of every experimental physicist or manufacturer of measuring devices.

From these considerations the following basic calculations for some
numerical values in ur theory become understandable.
We refer to considerations made by von Weizs\"acker \cite{cfw71} and
later by G\"ornitz \cite{goernitz88a} to estimate the number of urs
invested in particles.
As pointed out, the decision of an ur alternative
is thought to be a decision in position space.
Thus the simplest decision which can be made on a particle will be to
decide if it is localized in the ''left'' or the ''right'' half
of the cosmos and is therefore the decision of an ur alternative.
Now it is well known that the Compton wavelength $\lambda=\frac{h}{mc}$
gives a measure of the size of a massive particle.
Because of the empirical fact that the ratio between the cosmic radius
$R$ and the Compton wavelength of a proton is about
\be
\frac{R}{\lambda_p} = 10^{40}
\ee
the proton can be considered as containing
\be
n_p = 10^{40}
\ee
urs. From the same argument it follows that $n_e=10^{38}$
is the number of urs in an electron.
Now, how many urs are there in the universe?
For a size measurement of the order
\be
\label{unbestimmt}
\Delta x \simeq \frac{hc}{E}
\ee
a measuring particle with the energy $E$ is needed.
Since the main part of the total energy of the universe comes from
ponderable matter, i.e., from protons or nucleons (as we know today),
the volume $\lambda_p^3$ gives an approximation of an elementary cell
of volume in which the whole cosmos could in principle be divided
simultaneously.
Then it follows for the total number of urs in the universe
\be
N = \frac{R^3}{\lambda_p^3} \approx 10^{120} .
\ee
It was the first empirical test for ur theory that from this result the
correct number of nucleons in the world is given by
\be
z_p = \frac{N}{n_p} \approx 10^{80} .
\ee
To verify these results we could imagine a single ur as a wavefunction
expanded over the whole cosmic space, i.e., a wavefunction with minimal
localization. Thus from the uncertainty relation (\ref{unbestimmt})
it follows for the energy of a single ur
\be
\label{E_ur} E_o \simeq \frac{hc}{R} \approx 10^{-32} \ {\rm eV} .
\ee
Now this value is indeed compatible with the above results, because for
the total energy of the universe we get
\be
U = N \cdot E_o = z_p \cdot E_p \approx 10^{88} \ {\rm eV}
\ee
with the proton energy $E_p=1$ GeV.

On the basis of similar considerations it is possible to derive
$n_{ph}=N^{\frac{1}{4}} \approx 10^{30}$ for the number of urs in a
photon. With $z_{ph}=\frac{N}{n_{ph}} \approx 10^{90}$ for the number
of photons in the world one finds the correct empirically verified value
for the photon-baryon ratio $\frac{z_{ph}}{z_p} \approx 10^{10}$.

It seems reasonable to interpret these results as a proper confirmation
of the ur-theoretic estimates, because the correspondence of such large
numbers cannot be dimissed as a product of mere chance.
The quantum theory of urs as a quantum theory of information
originally has to deal with astronomical ur numbers, i.e.,
the {\em information contents of the physical objects in bit}.
Thus ur theory provides a natural way of motivating such enormous
physical numbers discussed, for example, by Eddington
\cite{eddington31} or Dirac \cite{dirac37}.\\

Since the overwhelming part of information is needed to represent
a physical object as localized in position space,
the existence of large bit numbers cannot be reproduced in standard physics.
Only when considering physical extremes, e.g.,
a particle passing over the event horizon of a black hole,
does the knowledge about its whole informational content get lost.
In this case ur theory explains a result found in black hole theory:
the difference of the Bekenstein-Hawking entropy \cite{bekenstein73}
for a particle of mass $m$ falling into a black hole of mass $M$ is
\be
\Delta S = 4\pi ( (M+m_p)^2 - M^2 ) = 8\pi M m_p .
\ee
Now there is a close connection between the concept of entropy
and information. As von Weizs\"acker \cite{cfw85} has pointed out,
entropy has to be understood as {\em potential information},
i.e., information which can be won if one is interested in the actual
microstate of a system. For this reason the entropy of a physical system
is of the order of the number of urs in it.
G\"ornitz \cite{goernitz86} has shown that for a proton falling
into a black hole the maximal loss of information, which yields
$M=M_{u}$ (mass of the universe), is exactly the ur-theoretic value
for the informational content of the proton given above
\be
\Delta S_{max} = 8 \pi M_{u} m_p
    \approx 10^{55} g \cdot 10^{-24} g \approx 10^{41} m_o^2
\ee
where $m_o=10^{-5} g$ is the Planck mass.
This again shows the important aspect in an information-theoretic
interpretation of ur theory that the overwhelming empirically possible
amount of information in an object is invested in its
spatiality and is therefore not taken into consideration in ordinary
physics or information theory. One could say that this information,
contributing to the possibility of localization of particles,
does not appear in standard physics because
it is hidden in the semantics of the concept of a particle,
which is presupposed in common physics.

\section{\mbox{Energy and Matter as Condensates of}
         \mbox{Information}}

\subsection{Vacuum Energy Density as a Density of Information}

From ur-theoretic considerations G\"ornitz
\cite{goernitz88b}, \cite{goernitz_etal89}
has derived the existence of a nonvanishing effective cosmological constant,
i.e., a time-dependent cosmological term which yields
\be
\Lambda(t) \sim \frac{1}{R^2(t)} \approx 10^{-120} \ {\rm cm}^{-2}.
\ee
Its numerical smallness, a problem in ordinary cosmology, in ur theory
appears as a natural result.
Since space has to be regarded as the representation of information,
a nonvanishing cosmological constant has to be understood as a
necessary consequence. The existence of $\Lambda \ne 0$ is often
regarded as indicating the ontological priority of space over matter.
This was the reason for Einstein to dismiss his own invention,
because the Mach principle, which does not allow the space to have
any physical effect, would be violated. But from
the ur-theoretic point of view this seems like begging the question.
Space is by no means ''empty'', it is at least filled up with urs.
Moreover, its structure as a global $\setS^3$ is a consequence of the
isomorphic structure of the abstract symmetry group of urs, i.e.,
space is the appearance of pure information in the world.
Apart from this further appearances of information like energy
and matter exist. Hence radiation and massive particles as well as the
vacuum energy density will be described by densitiy situations of urs,
i.e., of information.
In that sense energy and matter can be looked upon as condensates
of information in front of a background of urs representing the vacuum.

\subsection{Particle Physics in Minkowski Space}

In ur theory the global world model is $\setS^3$. But according to
Wigner the states of elementary particles can be considered as
representations of the Poincar\'e group and therefore the particle
concept is only defined in the approximation of a flat Minkowski space.
As Castell \cite{castell75a} pointed out, the complex conjugation in
(\ref{Q}) leads to an introduction of anti-urs and, as a consequence, to
the new symmetry group $SU(2,2)$ which is locally isomorphic to the
conformal group $SO(4,2)$.
The Poincar\'e group is a subgroup of $SO(4,2)$.

In order to build particle representations a quantization procedure
is needed to allow anihilation and creation of urs
in the tensor Fock space
\be
T^{(R)} = \bigoplus_n T^{(R)}_n
\ee
whereas $T^{(R)}_n$ is the tensor product space over an
$R$-dimensional complex vector space $V^R$ spanned by
the $R$ basis vectors of urs ($R=2$ as defined in
(\ref{tensorspace}) ) or of urs and anti-urs ($R=4$).
Now in ur theory a parabose quantization is used, i.e,
the most general commutation relations which are compatible
with the Heisenberg equations.
This generalization of statistics was first suggested
by Green \cite{green53}.
We use the following abbreviations
\be
\label{abbrev}
 \alpha_{rs}     = \frac{1}{2} \Big\{ a_{r}    , a_{s}     \Big\} , \qquad
 \alpha_{rs}^{+} = \frac{1}{2} \Big\{ a_{r}^{+}, a_{s}^{+} \Big\} , \qquad
 \tau_{rs}       = \frac{1}{2} \Big\{ a_{r}^{+}, a_{s}     \Big\}
\ee
and for the number operator
\be
 n_{r}           = \tau_{rr} - \frac{p}{2} , \qquad \qquad
 n               = \sum_{r} n_{r} .
\ee
Now parabose quantization is done, if the anihilation
and creation operators of urs $a_r$, $a_r^+$ ($r=1...R$)
fulfill the (trilinear!) Green commutation relations
\be
 \Big[ a_r , \tau_{st}   \Big] = \delta_{rs} \ a_t \ , \qquad
 \Big[ a_r , \alpha_{st} \Big] = \Big[ a_r^+ , \alpha^+_{st} \Big] = 0
\ee
These conditions can be satisfied if the operators $a_r$, $a_r^+$
are given in the Green decomposition
(the parameter $p$ is called parabose order)
\be
a_r   = \sum_{\alpha=1}^p b_r^{(\alpha)}  \ , \qquad \qquad
a_r^+ = \sum_{\alpha=1}^p b_r^{(\alpha)+}
\ee
with the following commutation relations for the
Green components $b_r^{(\alpha)}$, $b_r^{(\alpha)+}$
\be \ba{lllllllll}
   \left[ b_r^{(\alpha)},   b_s^{(\alpha)+} \right]  & = &
   \delta_{rs} \ ,                                   &   &
   \left[ b_r^{(\alpha)+},  b_s^{(\alpha)+} \right]  & = &
   \left[ b_r^{(\alpha) },  b_s^{(\alpha) } \right]  & = &  0 \ , \\ \\
   \left\{ b_r^{(\alpha)},  b_s^{(\beta)+}  \right\} & = &
   \left\{ b_r^{(\alpha)},  b_s^{(\beta) }  \right\} & = &
   \left\{ b_r^{(\alpha)+}, b_s^{(\beta)+}  \right\} & = & 0 &   &
   (\alpha \ne \beta) \ .
\ea \ee
A paraboson can be looked upon as an object which consists of $p$
bosonlike subobjects.
This can be seen by considering the theory of Young diagrams.
One diagram  is a frame of $n$ boxes arranged in rows and columns
in which the number of boxes per row does not increase downward.
The classes of equivalent irreducible representations of the
symmetric group $S_n$ can be illustrated by the Young diagrams.
For example, consider the diagrams for three objects
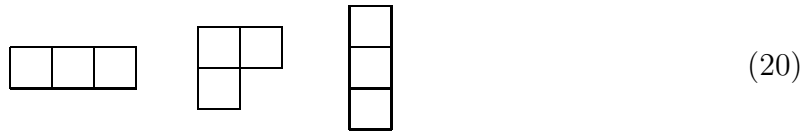
\be \label{young}
    \ba{|c|c|c|}
                                                 \cline{1-3}
     \phantom{8} & \phantom{8} & \phantom{8}  \\ \cline{1-3}
    \ea
    \qquad
    \ba{|c|c|c}
                                            \cline{1-2}
     \phantom{8} & \phantom{8} &         \\ \cline{1-2}
     \phantom{8} & \multicolumn{2}{c}{ } \\ \cline{1-1}
    \ea \!\!\!\!
    \qquad
    \ba{|c|}
                    \cline{1-1}
     \phantom{8} \\ \cline{1-1}
     \phantom{8} \\ \cline{1-1}
     \phantom{8} \\ \cline{1-1}
    \ea
\ee
A diagram in which the numbers $1 \ldots n$ are filled in obeying the rule
that they increase in each row from the left to the right and also in each
column from the top to the bottom is called a {\em standard tableau}.
The number $f_k$ of each type of tableau $k$ gives the number of the
irreducible representations of $S_n$ and also their dimension.
For example, the two possible tableaux
$\ba{|c|c|c}
                                  \cline{1-2}
     1 & 2 &                   \\ \cline{1-2}
     3 & \multicolumn{2}{c}{ } \\ \cline{1-1}
    \ea \!\!\!\! $
and
$\ba{|c|c|c}
                                  \cline{1-2}
     1 & 3 &                   \\ \cline{1-2}
     2 & \multicolumn{2}{c}{ } \\ \cline{1-1}
    \ea \!\!\!\! $
of the mixed-symmetric type ($k=2$) both correspond to two-dimensional
representations (i.e., $f_2=2$) of $S_3$.
This leads to the well-known formula
\be
     \sum_k f_k^2 = n!
\ee
A diagram in which the numbers do not decrease in each row and
increase in each column is called a {\em standard scheme}.
In $T^{(R)}_n$ every diagram defines $f_k$ irreducible
representations of the full linear group $GL(R)$. Each scheme
defines the basis vectors for these representations, e.g.,
a four-dimensional representation of $GL(2)$ in $T^{(2)}_3$
is defined by the diagram
$\ba{|c|c|c|}
                                                 \cline{1-3}
     \phantom{8} & \phantom{8} & \phantom{8}  \\ \cline{1-3}
  \ea$
and is given by the tensors
\begin{eqnarray*}
  \ba{|c|c|c|}
               \cline{1-3}
  1 & 1 & 1 \\ \cline{1-3}
  \ea & \quad & \ket{\phi_{111}} = \ket{111} , \\
  \ba{|c|c|c|}
               \cline{1-3}
  1 & 1 & 2 \\ \cline{1-3}
  \ea & \quad & \ket{\phi_{112}} = \ket{112} + \ket{121} + \ket{211} , \\
  \ba{|c|c|c|}
               \cline{1-3}
  1 & 2 & 2 \\ \cline{1-3}
  \ea & \quad & \ket{\phi_{122}} = \ket{122} + \ket{212} + \ket{221} , \\
  \ba{|c|c|c|}
               \cline{1-3}
  2 & 2 & 2 \\ \cline{1-3}
  \ea & \quad & \ket{\phi_{222}} = \ket{222} .
\end{eqnarray*}
Now, the two-dimensional representations are defined by the diagram
$\ba{|c|c|c}
                                            \cline{1-2}
     \phantom{8} & \phantom{8} &         \\ \cline{1-2}
     \phantom{8} & \multicolumn{2}{c}{ } \\ \cline{1-1}
 \ea \!\!\!\! $
and are given by two tensors in each case
\begin{eqnarray*}
    \ba{|c|c|c}
                                  \cline{1-2}
     1 & 1 &                   \\ \cline{1-2}
     2 & \multicolumn{2}{c}{ } \\ \cline{1-1}
    \ea & \quad & \biggl\{
        \displaystyle{ \ket{\psi_{112}} = 2 \cdot \ket{112}-\ket{211}-\ket{121}
                \atop  \ket{\psi_{211}} = 2 \cdot \ket{211}-\ket{112}-\ket{121}
                     } \\
    \ba{|c|c|c}
                                  \cline{1-2}
     1 & 2 &                   \\ \cline{1-2}
     2 & \multicolumn{2}{c}{ } \\ \cline{1-1}
    \ea & \quad & \biggl\{
        \displaystyle{ \ket{\psi_{122}} = -2 \cdot \ket{122}+\ket{221}+\ket{212}
                \atop  \ket{\psi_{221}} = -2 \cdot \ket{221}+\ket{122}+\ket{212}
                        }
\end{eqnarray*}
The parabose quantization procedure only admits tensors of urs which
correspond to young diagrams with maximal $p$ rows. For that reason
the number of rows cannot exceed $p=R$. All tensors of higher parabose
order are linearly dependent on the tensors of order $p \le R$.
It has been mentioned \cite{ohnuki+kamefuchi69} that this is not
sufficient to characterize parabose statistics complete.
The parabose procedure picks out only one tensor for every standard
scheme, i.e., the multiplicity of the irreducible subspaces of $S_n$ as
regards index permutations is always one.
For example, in the case of the scheme
$\ba{|c|c|c}
                                  \cline{1-2}
     1 & 1 &                   \\ \cline{1-2}
     2 & \multicolumn{2}{c}{ } \\ \cline{1-1}
    \ea \!\!\!\! $
for $p=2$ only the tensor
\be \ket{\psi_{121}} = \frac{1}{8} \left( a^+_1 a^+_2 a^+_1
       - \frac{1}{2} ( a^+_1 a^+_1 a^+_2 + a^+_2 a^+_1 a^+_1 ) \right)
       \ket{\Omega} = 2 \cdot \ket{121}-\ket{112}-\ket{211}
\ee
can be obtained which is a linear combination of $\ket{\psi_{112}}$ and
$\ket{\psi_{211}}$
\be
 \ket{\psi_{121}} = - \ket{\psi_{112}} - \ket{\psi_{211}}
\ee
whereas, for instance, $\ket{\phi_{112}}$ for $p=1$ is simply given by
$a^+_1 a^+_1 a^+_2 \ket{\Omega}$.
The other tensors of the higher dimensional representations can
be obtained by permutations of the indices, i.e., quantum numbers (place
permutations are not well defined). So it turns out as a consequence
that the parabose procedure corresponds exactly to the physically
distinguishable tensors in $T^{(R)}_n$, whereas tensors which can be
obtained by permutation of the ur indices are physically indistinguishable.

With parabose quantization one is able to represent Lie groups
in the Fock space of urs which is raised over the vacuum state defined by
\be \label{vac}
 b_r^{(\alpha)} \ \ket{\Omega} = 0  \quad \forall \ r,\alpha
\ee
and accordingly
\be
 a_r a_s^{+} \ \ket{\Omega} = p \ \delta_{rs} \ \ket{\Omega} \ .
\ee
The representation of the conformal group $SU(2,2)$ is given by
the 15 generators
\begin{eqnarray}
M_{12} &=& i/2 \ ( n_{1} - n_{2} + n_{3} - n_{4} )                , \nonumber \\
M_{13} &=& 1/2 \ (-\tau_{12} + \tau_{21} - \tau_{34} + \tau_{43} ), \nonumber \\
M_{23} &=& i/2 \ ( \tau_{12} + \tau_{21} + \tau_{34} + \tau_{43} ), \nonumber \\
M_{15} &=& i/2 \ ( \tau_{12} + \tau_{21} - \tau_{34} - \tau_{43} ), \nonumber \\
M_{25} &=& 1/2 \ ( \tau_{12} - \tau_{21} - \tau_{34} + \tau_{43} ), \nonumber \\
M_{35} &=& i/2 \ ( n_{1} - n_{2} - n_{3} + n_{4} )                , \nonumber \\
M_{46} &=& i/2 \ ( n + 2p )                                       , \nonumber \\
       & & \nonumber \\
N_{14} &=& i/2 \ ( \alpha_{13} + \alpha_{13}^{+} - \alpha_{24} - \alpha_{24}^{+} ), \nonumber \\
N_{24} &=& 1/2 \ (-\alpha_{13} + \alpha_{13}^{+} - \alpha_{24} + \alpha_{24}^{+} ), \nonumber \\
N_{34} &=& i/2 \ (-\alpha_{14} - \alpha_{14}^{+} - \alpha_{23} - \alpha_{23}^{+} ), \nonumber \\
N_{16} &=& 1/2 \ (-\alpha_{13} + \alpha_{13}^{+} + \alpha_{24} - \alpha_{24}^{+} ), \nonumber \\
N_{26} &=& i/2 \ (-\alpha_{13} - \alpha_{13}^{+} - \alpha_{24} - \alpha_{24}^{+} ), \nonumber \\
N_{36} &=& 1/2 \ ( \alpha_{14} - \alpha_{14}^{+} + \alpha_{23} - \alpha_{23}^{+} ), \nonumber \\
N_{45} &=& 1/2 \ ( \alpha_{14} - \alpha_{14}^{+} - \alpha_{23} + \alpha_{23}^{+} ), \nonumber \\
N_{56} &=& i/2 \ ( \alpha_{14} + \alpha_{14}^{+} - \alpha_{23} - \alpha_{23}^{+} ) .
\end{eqnarray}
The generators of the Poincar\'e group are then given by
\be
\begin{array}{ll}
M_{ik}                  & angular \ momenta, \\
N_{i4}                  & Lorentz \ boosts,  \\
P_{i} = M_{i5} + N_{i6} & momenta,           \\
P_{0} = N_{45} + M_{46} & energy \qquad (i=1,2,3).
\end{array}
\ee
To build particle representations, as a first step it is necessary
to describe a special vacuum state $\ket{\omega}$ which is
invariant under the Poincar\'e group. This is by no means a state
which is empty of urs (like $\ket{\Omega}$ in (\ref{vac}) is),
because in general the generators of the Poincar\'e
group change the number of urs. From the information-theoretic point
of view the Lorentz vacuum $\ket{\omega}$, a state in which no
particle exists, must be regarded as containing a lot of information,
i.e., its special Lorentzian space structure.
In \cite{goernitz_etal92b} the Lorentz vacuum is given by
\be
\ket{\omega} = \sum^{\infty}_{\mu=0} \sum^{\infty}_{\lambda=0}
    \frac{(-1)^{\mu+\lambda} i^{\mu-\lambda}}{\mu! \lambda!}
    \alpha^{+ \ \mu}_{14}
    \alpha^{+ \ \lambda}_{23} \ket{\Omega}
             = e^{ i (\alpha^+_{23} - \alpha^+_{14}) } \ket{\Omega} .
\ee
From this Lorentz vacuum state of urs it is now possible to build
particle states by applying the operators (\ref{abbrev}) on it.

For example, the representation of a massless spin-$0$ particle which
von Weizs\"acker calls a ''zeron'' for $p=1$ is given by
\be
\ket{\psi{\textstyle (s=0)}} =
\sum_{\mu=0}^{\infty} \frac{(i \epsilon)^{\mu}}{(\mu!)^2}
\ \alpha^{+ \ \mu}_{14} \ket{\omega}
\ee
whereas a massless spin-$\frac{1}{2}$ particle, i.e., a neutrino,
is only distinguished from the zeron by applying one additional ur
on it producing the spin in the $z$ direction
\be
\ket{\psi{\textstyle (s=\frac{1}{2})}} = a_1 \ \ket{\psi{\textstyle (s=0)}} .
\ee
These states fulfill the conditions
\be\ba{ccl}
 P_1  \ket{\psi} = 0 & \qquad & (P_0 - P_3) \ket{\psi} = 0 \\
 P_2  \ket{\psi} = 0 & \qquad & (P_0 + P_3) \ket{\psi}
 = i \epsilon \ \ket{\psi} \qquad p_0=p_3=\frac{\epsilon}{2}
\ea\ee

Further representations of massive particles with spin can be obtained in
the same manner. This was first done in \cite{goernitz_etal92b},
but investigations into these states and their correspondence
to the known types of fundamental particles like quarks and leptons
are still underway.
This is the actual ur-theoretic way to try to find the connection
between ur theory and the standard model of elementary particle physics.


\section{Cosmic Evolution as an Evolution of Information}

Finally, cosmic evolution should be briefly discussed from the ur
and information-theoretic point of view.
Evolution can in principle be regarded as a production of more and more
complex structures - in a first step the formation of elementary
particles, later on the formation of atoms, molecules, planets,
biological cells, animals, and so on.
In information-theoretic language these structures of different
complexity could be regarded as different semantic levels.
It therefore has to be an aim of ur theory to describe this evolution
as an evolution of information, where the formation of higher semantic
levels can be explained from the levels below. The elementary objects,
urs, represent the lowest semantic level, i.e., simple bits.
Now the change into the next level is equivalent to the introduction of
a new structural feature which allows the forming of classes of urs.
This forming of classes can then be regarded as a new semantic level
and exactly this is done by using parabose quantization in ur theory,
because the parabose parameter $p$ is an index for the different types
of urs.

In ur theory a method would be needed to describe the change
of semantic levels in a general way. This could presumable be done in
von Weizs\"acker's procedure of {\em multiple quantization} (in German,
{\em Mehrfache Quantelung}), which has to be regarded as an iteration
of complementarity logic \cite{cfw_etal58}.
This means that the components of an $n$-fold alternative correspond
to complex-valued truth variables.
The question arises of how this procedure can be connected
with the parabose quantization to reach higher levels.
The first three semantic levels so far seem to be: ur alternatives,
parabose tensors of urs, and elementary particles as described above.
Thus a straightforward procedure of multiple quantization
could be a mathematical way to explain evolution in the framework of
a quantum theory of information as an iteration of semantics.

\subsection*{Acknowledgments}

I thank Prof. M. Drieschner for his support and many discussions.
I also thank Prof. Th. G\"ornitz and Prof. C. F. von Weizs\"acker
for helpful and stimulating remarks on the manuscript.
I am grateful to the Graduiertenf\"orderung of the Ruhr-University Bochum
for financial support.

\end{document}